%Paper: 9108008
%From: COBI@slacvm.slac.stanford.edu
%Date: Tue, 20 Aug 1991 15:17 -0800 (PST)

\input phyzzx
\overfullrule=0pt
\def\p{\partial}
\def\bp{\bar\partial}
\def \l{\lambda}
\def \bl{\tilde\l}
\def \tw{\tilde w}
\def \t{\tilde}
\def \bphi{\tilde\phi}
\def \e{\epsilon}
\def\w{\omega}
\def\tw{\tilde W}

\def\W{W^{(n)}}
\def\Q{Q^{(n)}}
\def\R{R^{(n)}}
\def\G{G^{(n)}}

\vskip 2cm
%
% journal definitions
%
\def\cmp#1{{\it Comm. Math. Phys.} {\bf #1}}

\def\pl#1{{\it Phys. Lett.} {\bf #1B}}

\def\prd#1{{\it Phys. Rev.} {\bf D#1}}

\def\np#1{{\it Nucl. Phys.} {\bf B#1}}

\def\jmath#1{{\it J. Math. Phys.} {\bf #1}}
\def\mpl#1{{\it Mod. Phys. Lett.}{\bf A#1}}

\REF\wtym{E. Witten, \cmp {117} (1988) 353.}
\REF\wtrr{E. Witten, \np {340} (1990) 281.}
\REF\WD{R. Dijkgraaf, E. Witten, \np {342} (1990) 486.}
\REF\vv{ E. Verlinde, and H. Verlinde,
``A solution of two dimensional topological quantum gravity"

 Princeton preprint PUPT-1176.}

\REF\dvv{R. Dijkgraaf, E. Verlinde, and H. Verlinde,
\np {352} (1991) 59;
``Notes On Topological String THeory And 2D Quantum Gravity,'' Princeton
preprint PUPT-1217 (1990).}
\REF\ams{K. Aoki, D. Montano, J. Sonnenschein,\pl {247} (1990) 64.}
\REF\ms{D. Montano, J. Sonnenschein, \np {313} (1989) 258.}
\REF\lpw{J. Labastida, M. Pernici,  E. Witten, \np {310} (1988) 611.}

\REF\mss{D. Montano, J. Sonnenschein ,  \np {324} (1989) 348.}
\REF\js{J. Sonnenschein \prd {42} (1990) 2080.}

\REF\eguchi{T. Eguchi and S.-K. Yang, \mpl {5} (1990) 1693.}
%``$N=2$ Superconformal Models as
%Topological Field Theories,'' Tokyo preprint UT-564.}
\REF\li{K. Li, \np {354} (1991) 711, ibid. 725}
%`Topological Gravity With Minimal Matter,''
%Caltech preprint CALT-68-1662, ``Recursion Relations In Topological Gravity
%With Minimal Matter,'' Caltech preprint CALT-68-1670.}

\REF\sy{M. Spiegelglas and S. Yankielowicz, ``$G/G$ Topological Field Theory
by Cosetting,'' ``Fusion Rules As Amplitudes in $G/G$ Theories,''
preprints, to appear.}

\REF\dis{J. Distler, ``2-D Quantum Gravity, Topological Field Theory,
and the Multicritical Matrix Models,'' Princeton preprint PUPTHY-1161
(1990).}

\REF\wtsm{E. Witten \cmp {118} (1988) 411}
\REF\Walgebra{
E. Bergshoeff, C.N. Pope, L. J. Romans, E. Sezgin, X. Shen, and S. Stelle
\pl {243} (1990) 350, {\bf 245} (1990) 447, {\bf 256} (1990) 350;
``$w_\infty$ Gravity  and Super $w_\infty$ CERN-TH 5743/90;

E. Bergshoeff, M. Vasiliev and B. de Wit  \pl {256} (1991) 199 and references
therein.}
\REF\AMS{K. Aoki, D. Montano, J. Sonnenschein
``The role of Contact Algebra in the Multi Matrix Models"
Caltech preprint CALT-68-1676, to appear in {\it Mod. Phys. Lett.}}
\REF\FMS{ D. Friedan, E. Martinec and S. Shenkar \np {271} (1986) 93}
 \REF\GK{
  K. Gawedzki and A. Kupianen , Phys. Lett. 215B (1988) 119, Nucl. Phys.
     B320 (1989)649
     D. Karabali and H.J. Schnitzer Nucl. Phys. B329 (1990)649
     D. Karabali Gauged WZW Models and the Coset Construction of Conformal
     Field Theories", talk at the NATO ARW 18th Int. Conf. on Differential
     Geometric Methods in Theoretical Physics, Tahoe CA, 1989
     H.J. Schnitzer, Nucl. Phys. B324(1989)412
     K. Bardacki, E. Rabinovici and B. Saring, Nucl. Phys. 299(1988)157}
\REF\Bos {  J. Distler and Z. Qiu, Nucl. Phys.B336(1990)533
           D. Nemeschansky, Phys. Lett.B224(1989)121;
 Gerasimov, A. Marshakov, A. Morozov, A. Olshanetskii, S. Shatashvilli
 {\it Int. J. Mod. Phys.} A5(1990)2495}

 \REF\Nem{ D. Nemeschansky and S. Yankielowicz  1990 unpublished
      see also M. Spigelglas and S. Yankielowicz ref. [\sy]
      in which the U(1) case is discussed in details.}
\REF\pol{
   R. Budzynski and M. Spalinski, "on the chiral algebra of topological
 conformal field theories", TUM-TH-121/91 preprint.}
\REF\roger {R. Brooks ``Residual Symmetries in c=1 noncritical string
theory" MIT-CTP 1957 (1991).}
%%%%%%%%%%%%%%%%%%%%%%%%%%%%%%%%%%%%%%%%%%%%%%%%%%%%%%%%%%%%%%%
\rightline{TAUP-1898-91}
\date{August 1991}

\titlepage
\vskip 1cm
\title{Novel Symmetries of Topological Conformal Field  theories}
\author {J. Sonnenschein and S. Yankielowicz
\footnote{\dagger}{Work supported in part by the US-Israel Binational Science
Fundation and the Israel Accademy of Sciences}}
\address{ School of Physics and Astronomy\break
Beverly and Raymond Sacler \break
Department of Exact Sciences\break
Ramat Aviv Tel-Aviv, 69987, Israel}
\abstract
{ We show that various
 actions of  topological conformal theories that were
suggested recentely are particular cases of a general action. We prove
the invariance of these models under transformations generated by nilpotent
fermionic generators of arbitrary conformal dimension,
$\Q$ and  $\G$.
The later are shown to be the $n^{th}$ covariant derivative with respect
to ``flat abelian gauge field" of the fermionic fields of those models.
We derive the bosonic counterparts $\W$  and $\R$
which together with  $\Q$ and
$\G$ form
 a special $N=2$ super
$W_\infty$ algebra. The algebraic structure is discussed and it is shown
that it generalizes the so called ``topological algebra".}

\section{ Introduction}
\par
Two dimensional gravity and non-critical string theories provide an
interesting and useful arena for the study of topological quantum field
theories(TQFT's)\refmark\wtym.
In the opposite direction, the general covariant formulation
%The idea that  general covariant formulation,
%which is
%equivalent to a system after the integration of all metrics,
provides an
effective tool to calculate  ``correlation functions"  in the
form of  powerful recursion relations\refmark{\wtrr -\ams}.
Several different starting points for  topological conformal field theories
(TCFT's) were proposed. It was
realized that ``pure gravity",\refmark{\ms,\lpw}
 flat two dimensional gauge connection\refmark{\mss,\js}, twisted
$N=2$ superconformal theories\refmark{\eguchi,\li},
the ${G\over G}$ construction\refmark\sy and topological
sigma models\refmark{\wtsm,\ms,\WD}
  were all examples of TCFT's.

 In this work we elaborate on the equivalence between the various models
and suggest a general framework to analyze all of them.
We  show that all these models are invariant under transformations
generated by infinitely many
bosonic and fermionic generators of arbitrary  integer conformal dimension.
The fermionic generators  are nilpotent. They
have the structure of higher
order covariant derivative with respect to a flat gauge connection of
the fermionic fields of the models.
This symmetry may be referred to as ``$N=2$ super $ W_\infty$" symmetry.
We discuss the algebraic structure of those symmetries,
which generalize the minimal
``topological algebra"
of ref. [\dvv] and
present several useful OPEs'.
Infinite towers of bosonic symmetries as well as supersymmetries were discussed
in the past.\refmark\Walgebra\ Both the fermionic and bosonic symmetry
generators discussed here are different though possibly not unrelated to those
in the literature.

  \par The paper is organized as follows
In section 2 we describe the various formulations of
conformal topological quantum field theories. We start with Einstein's action
of
pure gravity, continue with the twisted $N=2$ models with zero and non-zero
background charge,  theories of flat gauge connections, the $G\over G$
construction and finely the general BRST invariant( $h,\ 1-h$) systems. The
equivalence of several of these formulations is discussed in section 3.
Section 4 is devoted
to the symmetries of the TCFTs' generated by a set of infinitely many bosonic
$W_\infty$ as well as fermionic $Q_\infty$ and $G_\infty$ generators. The
transformations of the various fields are written down. It is shown that the
 $n^{th}$ generator in each sector can be expressed in terms of an $n^{th}$
covariant derivative with respect to a ``flat gauge connection".  The structure
of this infinite topological algebra is investigated  in section 5.
In section 6 we summarize the results and comment briefly on the
possible implications.
Some technical details are presented in the appendices. In Appendix A
we prove the
invariance under $\Q$, construct the bosonic operator  $\R$ and
exhibit the corresponding
transformation  properties of the various fields. In Appendix B the calculation
of the anomaly term     in the OPE $\Q G^{(m)}$ is explained. Appendix C
presents the OPE of $W^{(1)}$ with the rest of the operators.

%In section 5 we
%make use of the infinitely many Nilpotent fermionic symmetries to show that
%the Hilbert space of all these theories is composed only from the zero modes
%of the various fields. In the last section we discuss the relation to matrix
%models.

 \section{ Actions for Topological Conformal Field
Theories}
The most obvious TQFT in two dimensions is pure Einstein's gravity ( with no
cosmological constant). In refs. [\ms , \lpw]
 it was realized that this action has in
addition to the usual scale and reparametrization invariance a local symmetry
 referred to as the ``topological symmetry". The BRST quantized action of
this theory  was found out to be\refmark{\ms,\lpw}:
$$ S_{TG}=\int d^2 z[( b\bp c + \beta\bp\gamma +cc ) +
\pi(\p\bp\varphi-\hat R^{(2)})
+ \t \psi \p\bp\psi] \eqn\mishaab$$
where $(b, c)$ are the spin (2,-1) reparametrization ghosts, $(\beta , \gamma)$
are commuting ghosts with the same spins, and $(\pi, \varphi)$ ,
$(\t \psi , \psi)$ are Grassmann even and odd scalars.
This action was derived via two stages of gauge fixing . In the first
the gauge condition $R^{(2)}=\hat R^{(2)}$ was imposed to fix
the ``topological symmetry".
 In the second stage the ordinary reparametrization was fixed
$ ds^2 =e^{2\varphi}
dz\bar dz$ as well as a ghost symmetry.
A different derivation of this action was given in ref. [\dis]
where it  was shown  to correspond
to a $C=-2$
matter theory coupled to a Liouville mode.

An alternative  formulation of pure gravity was written down in terms of flat
$SL(2,R)$ gauge connections\refmark\mss.  This construction was a special
case of
topological flat
gauge connection (TFC)\refmark\js
associated with  the group $G$. The action of this system
is described in the ``semi-classical limit" which is exact by
$$S_{FC}= \int d^2 z Tr[( A\bp \t B +\Psi\bp \t \Psi + c.c) +\p \phi\bp\t \phi
+ \p c \bp\t c]\eqn\mishaac$$
 where all the fields are in the adjoint representation of the group.
 $A$ is the gauge connection, $\Psi$ is a spin 1 ghost field related
to the gauge fixing of the topological symmetry, $(\t c, c)$ are the usual
ghost and anti-ghost of the non-Abelian symmetry
and $\t \phi, \phi$ are Grassmann even ``ghost for ghosts".
A detailed discussion of this model can be found in refs. [\mss, \js].
This scenario was invoked also to describe world-sheet supergravity using the
graded Lie group $OSP(2|1)$\refmark\AMS.
In that case the gauge connection was decomposed
as follows: $ A =e^a J_a + \w J_3 +\chi^\pm J_\pm$ were $J_a\ ( a=1,2), J_3
$ and $J_\pm$ are the generators of $OSP(2|1)$ and $e^a,\ \w$ and $\chi^\pm$
where interpreted as the zweibein, spin connection and the gravitino
respectively.
Similar construction for other interesting groups like $SL(3,R)$ were worked
out in ref. [\li].

Yet another formulation of topological theories
was found by twisting the $N=2$ superconformal field
theory\refmark{\eguchi,\li}.
In the  case of the minimal $N=2$ models the action takes the form of
 $$ S_{N=2}=\int d^2 z \p\phi\bp\bphi
+i\alpha\sqrt{g} R^{(2)}\phi +(\l\bar\p\bl
+c.c)\eqn\mishaa$$
where $\phi, \bphi$ are commuting scalars, $\l , \bl $ are  $(1,0)$ anti-
commuting ghosts, $ R^{(2)}$ is the world sheet curvature of the background
metric and $\alpha$ is a parameter of the theory.

A different description of a
TQFT which links it to some group $G$ was found by
extrapolating the ${G\over H} $ construction to the case of $H=G$\refmark\sy.
  It is well known that ${G\over H}$ coset models can be described in terms of
a
 WZW model based on a group $G$ where (an anomaly free) subgroup $H$
is gauged\refmark\GK.
 The gauging amounts essentially to setting the $H$-currents to zero. Hence
 for the case $H=G$ only the $G$-zero modes survive. In this case the system
 is equivalent to three decoupled systems i.e. $G$-WZW model at level $k$,
 $G$-WZW at negative level $-(k+c_G)$ and a free (1,0)-(b,c) system in the
 adjoint representation. Upon bosonization\refmark\Bos
  (assuming $G=SU(N)$ ) one can
 recast the action into a sum of the following terms.\refmark\Nem
  One term has the
 form of eqn.(3) where $\phi$ and $\t \phi$ are associated
with the ``hypercharge"
 currents of the two $SU(N)$-WZW models. (The scalars combine nicely into
 one complex boson). The ghost system in this term is the $(1,0) (b,c)$
 system associated with the hypercharge direction. The rest of the terms are
 $N^2-2$ free(1,0)-$(b,c)+(\beta,\gamma)$ systems each per one of the extra
 generators of $SU(N)$. Clearly one can use the bosonization formulas to
 recast this form into other equivalent forms.
 This structure can be modified by introducing appropriate background
 charges which do not change the value of the total central charge
$c_{tot}=0.$

 Older members of the family  are the topological sigma models (TSM)
\refmark{\wtsm,\ms,\WD,\dvv}. These
models describe a special sector of the maps from  two dimensional world
sheet into some target manifolds. If the target manifold is taken to be flat
then the expression of the corresponding action is given by:
 $$ S_{TSM}= \int d^2 z \eta_{\mu\nu}[ \p X^\mu \bp X^\nu +( \psi^\mu \bp
\t \psi^\nu  + cc )], \eqn\mishaad$$
$X^\mu$ is the target space coordinate, $\psi^\mu , \t \psi^\nu$ are
world-sheet
$(1,0)$ system.
Just as in the case of TFC, one can deduce this  quantum action via a BRST
gauge fixing of a topological symmetry.
\par  The later models as well as the previous ones
exhibit an important relation to
 non-critical string theories and their matrix models counterparts
once they are coupled to the topological two
dimensional gravity\refmark{\WD,\dvv} of eqn. \mishaab\ .
For instance when a TSM is coupled to TG to produce a ``topological
string" model,\refmark{\mss,\dvv} the corresponding action is the sum of
the actions given in eqn. \mishaab\ and \mishaad .
\par An interesting question is to what extent are ordinary critical
and non-critical string theories, both bosonic and supersymmetric, a
special case of TCFTs'.\refmark\dvv Apart from a comment in the last
section, we do not consider here string theories as TCFTs'.
\section { A ``unified" picture}
 \par Two questions are now in order: (i) can one ``unify"  the actions
of the models presented in the last section
 and (ii) are there other topological models. A straightforward
observation is that for $R^{(2)}=0$
all the actions described in the previous section are special cases
of the following general action:
%the following action provides an answer to those  questions:
$$ S =\int d^2 z \sum_i[\Phi^{(h_i)}\bp\tilde\Phi^{(1-h_i)}
+\Psi^{(h_i)}\bp\tilde\Psi^{(1-h_i)}+cc] \eqn\mishabb$$
where  $\Phi^{(h_i)}$,  $\Psi^{(h_i)}$
are commuting and anti-commuting fields of
dimension $(h_i)$ and similarly  for the dimension $(1-h_i)$ fields
$\tilde\Phi^{(1-h_i)}$ and $\tilde\Psi^{(1-h_i)}$.
For the terms involving a pair of scalars ( commuting or anti-commuting)
the passage to the  form of the above equation involves a simple redefinition
which amounts to rewrite them is a first order form.
For instance for eqn. \mishaa\ we rewrite $\p\phi\bar\p\bphi=W\bp\bphi$ with
$W=\p\phi$.
Since systems which are the same apart from their Grasmannian nature,
have conformal anomaly which differ by a sign,
it is obvious that the total
 conformal anomaly vanishes  $c=\sum_i (c_i-c_i)=0$.
One can
reformulate \mishabb\ as an exact form under fermionic operators $Q$ and $G$
of dimensions
zero and one  respectively.
$$S=\int d^2z \sum_i [Q(\Psi^{(h_i)}\bp\tilde\Phi^{(1-h_i)}) +cc]
 = \int d^2 z \sum_i
[G(\t\Psi^{(1-h_i)}\bp\check\Phi^{(h_i-1)}) +cc]
\eqn\mishaf$$
where $\Phi^{(h_i)}=\p \check\Phi^{(h_i-1)}$.
 The $Q$ and $G$ transformations of the various fields are given by

$$\eqalign{\delta^Q\Psi^{(h_i)} &=\e\Phi^{(h_i)}
\qquad\delta^Q\t\Phi^{(1-h_i)} =-\e\t\Psi^{(1-h_i)}\cr
\delta^G\t\Psi^{(1-h_i)} &=\e\p\t\Phi^{(1-h_i)}
\qquad\delta^G\Phi^{(h_i)} =-\e\p\Psi^{(h_i)}\cr}\eqn\mishaff$$
The fact that the action is exact under a zero dimension fermionic symmetry
 hints of   the  possibility
to interpret the action as a BRST gauge fixed action.
This  interpretation follows the original TQFTs' namely, that the
``classical" action is zero and the `` quantum action" is derived by
gauge fixing of a ``topological symmetry". One can take
${\cal L}_{classical}(\Psi^{(h_i)})=0$
which is invariant under the ``topological symmetry" $\delta
\Psi^{(h_i)} =\e^{(h_i)}(z,\bar z)$ or
${\cal L}_{classical}(\t\Phi^{(1-h_i)})=0$
which is invariant under the ``topological symmetry" $\delta
\t\Phi^{(1-h_i)} =\e^{(1-h_i)}(z,\bar z)$. Replacing the
parameters of transformations with ghost fields $\Phi^{(h_i)}$ for the first
formulation and $\t\Psi^{(1-h_i)}$ for the second, imposing holomorphicity of
the original fields as the gauge condition $\bp\Psi^{(h_i)}=0$ or
$\bp\t\Phi^{(h_i)}=0$, and using the ``BRST" transformations of eqn.
\mishaff\ one gets the action \mishabb\ .
Notice that this prescription is different from the BRST gauge fixing
that was
 applied for the cases of pure gravity\refmark\ms, TFC\refmark\mss
 and TSM\refmark\ms.
It is thus apparent that various different starting points for
TCFT lead to the same theory. This point will be further discussed in the
last section.

So far we considered only the case of $R^{(2)}=0$. For the twisted $N=2$
action it is equivalent to taking $\alpha=0$ which is
the semiclassical
limit of this action since $k=({1\over \alpha^2}-2)\rightarrow \infty$.
As we show in the next section, for pure imaginary $\alpha$, namely, negative
$k$, one can generalize the construction  by redefining
$\t\phi\rightarrow\hat\phi=\t\phi-i\alpha\phi$.

\section { The symmetries }
By  definition,
all  the ``physical observables"  of a TQFT
are invariant under an arbitrary variation of the metric of the
underlying manifold. (The notion of physical observables refers to correlation
functions of  products of operators which are scalars and gauge invariant with
respect to any local symmetry in the system.)
 This implies that the energy momentum tensor
can be expressed as an exact operator under a nilpotent fermionic symmetry
$$T_{\alpha\beta} = \{Q, G_{\alpha\beta} \}.\eqn\mishad $$
It is straightforward to check that eqn. \mishad\  guarantees the metric
independence.\refmark\wtsm\
Moreover, it is easy to see that in fact the TQFT actions given in the previous
section are all exact under the fermionic symmetry. This is obviously the
situation for the TG,
TFC and TSM models since the quantum action by construction is BRST
exact as well as for any other model following eqn. \mishaf\ .

  A topological conformal field theory (TCFT) is characterized by
the fact that the trace of the classical energy  momentum vanishes.
All the TQFT models presented in the previous section share this property.
 In these cases $T_{\alpha\beta}$
 as well as $G_{\alpha\beta}$, the BRST current $Q_\alpha$ and the ghost
number current $J_\alpha$ can be split into their
holomorphic and antiholomorphic
parts. Hence  one gets the following relations\refmark\dvv
which reflect the BRST multiplet structure
 $$T(z) =\{ Q,G(z)\} \ \  \ \  \ Q(z) =-[ Q,J(z)]\eqn\mishab $$
By Laurant expansion of  these operators one finds using Jacobi identities
the TCFT algebra. This algebra together with its generalization  will be
presented in the next section.

Next we analyze the symmetries of the TCFT.
Let us  first discuss the symmetries generated by $J,\ T,\ Q,\ $ and $G$. To
simplify the notation we choose to  demonstrate all the features in the twisted
$N=2$ model eqn. \mishaa\  with $\alpha=0$ or $R^{(2)}=0$.
 Later we explain how to generalize it
to the case of non-flat world sheet and to the
other models included in   the general form of the action given in eqn.
\mishabb\ .

The following transformations of the fields leave the  action  invariant.
$$\eqalign{ \delta^J\l =-\e\l \qquad &\delta^J\bl =\e\bl
\qquad\delta^J\phi =-\e \qquad \delta^J\bphi =\e\cr
\delta^Q\l =\epsilon\p\phi \qquad &\delta^Q \bphi =-\epsilon\bl\cr
 \delta^G\bl =-\epsilon\p\bphi \qquad &\delta^G \phi =\epsilon\l \cr
\delta^T\l =(\p\epsilon\l +\e\p\l) \qquad &
\delta^T\bl = \e\p\bl \qquad
\delta^T\phi = \e\p\phi \qquad
\delta^T\bphi = \e\p\bphi \cr}
\eqn\mishbb$$
The parameters of transformation $\e$  are holomorphic function $\e=\e(z)$.
For the $J$ and $Q$ transformations $\e$ has dimension zero, for  $T$ and $G$
 dimension one and for $Q$ and $G$ they are Grassmanian variables.
Obviously the action is invariant also under similar transformations generated
by the anti-holomorphic counterparts,
$\bar J, \bar T, \bar Q, \bar G $. From here on we discuss only the
holomorphic transformations.
Notice that unlike usual BRST transformations where the
parameter of transformation $\e$ is a global parameter, here $\e=\e(z)$
also for the fermionic symmetries generated by $Q$ and $G$.
Hence they generate an infinite dimensional algebra.
 Using the OPE of the basic fields
$$\phi(z)\bphi(\w)= -log(z-\w) \qquad  \l(z)\bl(\w) ={1\over
z-\w}\eqn\mishbc$$
it is straightforward to extract the  currents that generate   the
above transformations:
 $$\eqalign{ J=-(\l\bl + a\p\phi -\t a\p\bphi) \qquad
&T = -(\p\phi\p\bphi +\l\p\bl
+a\p^2\phi)\cr
 Q=\bl\p\phi +\t a\p\bl\qquad &G=-(\l\p\bphi+ a\p\l)\cr} \eqn\mishbd$$
Note that the terms proportional to $a$ and $\t a$
 are total derivatives
so they do not contribute to the corresponding charges and therefore cannot
be determined from the classical  transformations alone.
Even for parameters of transformations which are not global but rather are
holomorphic functions,
 in which case the total derivative terms do contribute to the
transformations, they cannot be determined.
 Hence one can generally
multiply each of them with an arbitrary parameter .
However, imposing the relations of eqn. \mishab\ reduces the
number of parameters
from five to two $ a$ and $\t a$ as stated in eqn. \mishbd\ .
These parameters will
play a role in the corresponding algebra as will be  discussed in the next
section.
  In fact there are some additional relations among the symmetry generators
 $$\t T(z) =\{ G,Q(z)\} \ \  \ \  \ G(z) =-[ G,\t J(z)]\eqn\mishabb $$
which are all summarized in the
following diagram:
$$\eqalign{ \t T, & T\cr Q \hskip 2 cm
&\hskip 2 cm G\cr  -J,& -\t J\cr}\eqn\mishbe$$
where $A--^B\rightarrow C$ denotes acting with a charge $B$ on a current $A(z)$
to generate a current $C(z)$.
The currents  $\t T$,  $T$ and $\t J$, $J$   correspond to the same symmetry
transformation and are
related to one another by $\phi\leftrightarrow -\t \phi$
and $\l\leftrightarrow\bl$.
%generator
%and differ from each other by the fact that the total  derivative term is of
% $\t \phi$ and $\phi$ respectively.
\par We wish now to address the question of whether the transformations of eqn.
\mishbb\  exhaust the symmetries of the TCFT models.
In what follows we consider only compact Riemann surfaces so the invariance
of the action will be checked always up to total derivatives.
The answer to this question is definitely no. The arsenal of symmetries is much
richer. There are in fact three types of symmetry transformations: (i) bosonic
or fermionic  transformations which involve only the commuting or the
anticommuting parts of the action like $\delta^J$ for $a=\t a =0$
(ii) bosonic symmetries acting
on both sectors like $\delta^T$ and (iii) fermionic symmetries  mixing
the two sectors like $\delta^Q$ and $\delta^G$. Before dwelling into the second
and third types let us write the most general invariance of each of the sectors
separately. Let us look for instance on the bosonic sector.  This part of the
action is invariant under the separate transformation of $\phi$ and $\bphi$ as
follows: $$\delta \phi= \e\p_{\tw}\tilde F(\tw)\qquad
 \delta \bphi= \e\p_{W}F(W)\eqn\mishbg$$
where $W=\p\phi$, $F(W)$ is a general function of $W$, $\p_W F(W)$ is
its derivative with respect to $W$ and similarly for the fields with
$\bphi $.  In particular any polynomials of $W$ and $\tw$ for $F$ and $\t F$
will do the job.
%In addition one has the invariance under the simultaneous
%transformations of $\phi$ and $\bphi$. .....
Symmetries which leave the fermionic sector invariant are for example those
which are generated by $\l$ (or $\bl$) $\delta\bl = \e$ ($\delta\l =\e$).
 \par Next we want to check whether there are
generalizations of the fermionic symmetries generated by  $Q$ and $G$. One
finds
that the following currents generate such symmetries.
$$\Q=D^n \bl\qquad \G=\tilde D^n \l \eqn\mishbm$$
where the covariant derivatives are  $D=\p+W=\p +\p\phi$ and  $\tilde
D=\p-\tw = \p -\p\bphi$ and $D^n$ is the $n^{th}$ power of $D$.
The fermionic generators $Q$ and $G$ are the special of $n=1$, $Q=Q^{(1)}$ and
$G=G^{(1)}$ with $\t a=- a =1$.
{}From the corresponding
OPEs' one gets that $\t a =-a$.
One can generalize the covariant derivative given above to incorporate $a\ne 1$
as in eqn. \mishbd\ and still maintain the structure of eqn. \mishbm\ by the
following redefinitions
$\l\rightarrow \l'=\t a \l\quad W\rightarrow W'={1\over\t a} W$
and the same for $\bl$ and $\t W$. Alternatively one can only redefine $\l$
and $\bl$ and      take ${1\over \t a}$ as the charge in the covariant
derivative.
 One can view this covariant derivative as if its source
is   an abelian gauge field which is taken to be a pure gauge, namely, zero
field
strength or   flat gauge connection.
%Is it related to flat connection of the TFC formulation?
We denote the set of infinitely many symmetry generators $Q^{(n)}$ as
$Q_\infty$
and $G^{(n)}$ as $G_\infty$.
It is obvious from eqn. \mishbm\  that the $G_\infty$ and the corresponding
transformation laws are related to those of $Q_\infty$ by the replacement
$\phi\rightarrow -\bphi$ and $\bl\rightarrow \l$. We thus describe here only
the  $Q_\infty$ symmetries.  Under the later only $\l$ and $\bphi$ transform
as follows:
$$\delta^{Q^{(n)}}\l =  D_-^{(n)} \e \qquad
\delta^{Q^{(n)}}\bphi =
-\sum_{i=0}^{n-1} D_-^{i}\e D^{(n-1-i)}\bl \eqn\mishbn$$
where $D_- =-\p +W$ such that $(DA)B-A(D_-B)=\p(AB)$.  The
parameter of transformation $\e$ has
 conformal dimensions $-(n-1)$.
To be specific here are the global transformations generated by the first three
lowest generators (omitting the parameter of transformations)
$$\eqalign{\delta^{Q^{(1)}}\l= W  &\qquad \delta^{Q^{(1)}}\bphi= -\bl \cr
\delta^{Q^{(2)}}\l =W^2-\p W &\qquad \delta^{Q^{(2)}}\bphi= -(\p\bl +2W\bl) \cr
\delta^{Q^{(3)}}\l = W^3-3 W\p W +\p^2 W &\qquad \delta^{Q^{(3)}}\bphi=
-(\p^2\bl + 3W\p\bl +3 W^2\bl).\cr} \eqn\mishbbn$$

It is straightforward to check that these transformations leave the action
invariant. In  appendix A we
 show that the general transformations eqn.  \mishbm\  are indeed symmetry
transformations.
\par So far  have we discussed the fermionic symmetries, now to complete the
generalization of the relation given by  eqn.
 \mishbe\  we define a double set of infinite bosonic operators as follows
$$ W^{(n)}(\w)={1\over 2\pi i}\oint_\w dz Q(z)G^{(n-1)}(\w)  \qquad
 \t W^{(n)}(\w)={1\over 2\pi i}\oint_\w dz G(z)Q^{(n-1)}(\w)   \eqn\mishbo$$
Using the OPE's of eqn. \mishbc\ we find for the $W_\infty $
$$W^{(n)} = \t D^{(n-1)} \hat W^{(1)} +G^{(n-1)}\bl \eqn\mishbp$$
where $\hat W^{(1)}= W-\l\bl$ and
$\t D W =(\p-\t W)W,\quad \t D \t \l =
\p\t \l $ . For some applications it is convenient to express $\W$ as follows:
$$W^{(n)} = \t D^{(n-2)} W^{(2)} -\sum_{k=0}^{n-2}(^{n-1}_k) G^{(k)}
\p^{n-1-k}\bl \eqn\mishbp$$
where  $G^{(0)} =\l$.
For $\t W^{(n)}$ we interchange $W$ with $-\t W$ and $\l$ with
$\t \l$. The expressions for the lowest $\W$ are
$$\eqalign{W^{(1)}&= W+\t W-\l\bl\cr
W^{(2)}&= \p W -W\t W -\l\p\bl\cr
W^{(3)}&= \p^2 W-\p (W\t W)-\t W\p W+ \t W^2 W -\l\p^2\bl -2\p\l\p\bl
+2\t W \l\p\bl \cr}\eqn\mishbpp$$
$ W^{(1)}$ is not determined by eqn. \mishbo\  . Its form is dictated by
the algebra of  the  generators.  Again it is easy to check that
$W^{(1)}=-J$ and $W^{(2)}=T$ with $\t a=- a=1$.
The invariance under the $\W$ transformations follows from that of $\G$:
$$\delta^{\W} S =[\e\W ,S]=[\e\{ Q, \G\} ,S] =\{ Q,[\G,S]\} +\{\e \G [Q,S]\}
=0\eqn\mishbppp$$
since $[Q,S]=[\G S]=0$. In the same manner $\Q$ invariance implies that
of $\t \W$. For completeness we write now the transformations
under the $W_\infty$ symmetry of
the various fields
$$\eqalign{&\delta^{W^{(n)}}\l =  (\t D_-^{(n-1)} \e)\l -\e \t D^{(n-1)}\l
\qquad
\delta^{W^{(n)}}\bl =  \t D_-^{(n-1)}( \e\bl) - (\t D_-^{(n-1)}\e)\l \cr
&\delta^{W^{(n)}}\bphi = -\t D_-^{(n-1)}\e \qquad  \delta^{W^{(n)}}\phi=
-\sum_{i=0}^{n-2} \t D_-^{i}\e D^{(n-2-i)} \hat W^{(1)} -\t D_-^i (\e \bl)
\t D^{(n-2-i)}\l \cr}
 \eqn\mishbr$$

where $\t D_-\e =-(\p +\t W)\e$.
How can we generalize
 the diagram of
eqn. \mishbe\ ? One is tempted to think that there is the same structure
also for the $n^{th}$ level. By definition
 $\W$ and $\t \W$ are created by applying $Q$ and $G$
on $\G$ and $\Q$ respectively,
however $\Q$ and $\G$ are not derivable from $W^{(n-1)}$
and $\t W^{(n-1)}$ by acting with  $Q$ and $G$.
One has to modify $\W$ and $\t \W $ in the following way to generate from
them $\Q$ and $\G$. First note that
$${1\over 2\pi i}\oint_\w dz Q(z)\t W^{(n+1)}(\w)= [\{Q,G\},\Q] +[\{Q,\Q\},G]
=[W^{(2)},\Q]=\p \Q. \eqn\mishbrr$$
Now it is easy to see that
if one adds $\l \Q $ to $\t W^{(n+1)}$ one gets
(see Appendix A):
$$ {1\over 2\pi i}\oint_\w dz Q(z)[\t W^{(n+1)}+\l\Q] (\w)=
{1\over 2\pi i}\oint_\w dz Q(z)\t R^{(n+1)} (\w)
=\p \Q+W\Q =Q^{(n+1)}. \eqn\mishbrr$$
and similarly for $G^{(n+1)}$.
The explicit expressions for $\R$ and $\t \R$ are
$$ \R= \t D^{(n-1)}\hat W^{(1)}\quad \t\R=  D^{(n-1)}\t {\hat W^{(1)}}.\eqn
\mishbbrr$$
The diagram of eqn.\mishbe\ takes now the split form
$$\eqalign{ \t W^{(n)},  W^{(n)}\hskip 2.7 cm
& \hskip 1.5 cm                    \cr
 Q^{(n-1)} \hskip 1.5 cm
\hskip 1.5 cm G^{(n-1)} \hskip 1cm  & \hskip 1 cm
 Q^{(n-1)} \hskip 1.5 cm
\hskip 1.5 cm G^{(n-1)}\cr
& \hskip 2.3 cm  \t R^{(n-1)},   R^{(n-1)}\cr}\eqn\mishbee$$

%The derivation of these relations are given in Appendix A.
Are the $\R$ and $\t \R$ generators of symmetries?
It is easy to see that $\delta^{\R} S $ is closed under $Q$.
$$[\Q,S] =[[Q,\R],S] = [Q,[\R,S]] -[\R,[Q,S]]=0.\eqn\mishbbee$$
It turns out, as we show in
Appendix A, that $\delta^{\R} S= \int d^z \bp(\e\l D^n\bl) =0$.
The transformations of the various fields under $\R$ are
also written down in the appendix.

\par The next task is to show that all these fermionic and bosonic symmetries
are in fact shared by all the models of the previous section.
To prove this  we first treat the general case of eqn.
\mishab\  and then we consider
the case of eqn.\mishaa\  for $R^{(2)}\ne 0$ and $\alpha\ne 0$.
The action \mishab\ is clearly a sum of decoupled actions, ( as long as
it is not coupled to TG)
so we can separate the symmetry generators for each separate
part
$Q_i^{(n)}$,  $G_i^{(n)}$  and $W_i^{(n)}$.
To construct the generators in a form similar to eqns.
\mishbm\ and \mishbp\   we need dimension one
fields as connections in the covariant derivatives.
For this purpose one can ``bosonize"\refmark\FMS
the bosonic system in eqn. \mishab\
in the following way
$$ \eqalign{
& \int d^2 z \sum_i[\Phi^{(h_i)}\bp\tilde\Phi^{(1-h_i)}]=\cr
&{1\over 2} \int d^2 z \sum_i [\p\rho_i\bp\rho_i -{1\over 4} Q_i
\sqrt{g} R^{(2)}\rho_i +2\eta_i\bp\xi_i]\cr}\eqn\mishbss$$
where
$\Phi^{(h_i)}(z)=e^{\rho_i(z)}\p\xi_i$ and
$\tilde\Phi^{(1-h_i)}=e^{-\rho_i(z)}\eta_i$ and $Q_i=-(1-2h_i)$.
Setting now $R^{(2)}=0$ we take $\p\rho_i$ as the connection of the
following covariant derivatives
$D_i  =\p +\p\rho_i$ and
$\t D_i  =\p -\p\rho_i$.
The expression for the symmetry
generators are thus
$$\eqalign{\Q_i&=D_i^n \t \Psi^{(1-h_i)}\qquad \G_i=\tilde D_i^n \Psi^{(h_i)}
\cr
W_i^{(n)} &= \t D_i^{(n-1)} \hat W_i^{(1)} +G^{(n-1)}\t \Psi^{(1-h_i)} \cr
&=
 \t D_i^{(n-2)} W_i^{(2)}
-\sum_{k=1}^{n-1}(^{n-1}_k) G_i^{(k)}
\p^{n-1-k}\t \Psi^{(1-h_i)}\cr} \eqn\mishbs$$
%-(\t D_i^{(n-1)}\Psi^{(h_i)} )\t \Psi^{(h_i-1)}\cr}
%\eqn\mishbs$$
 where now $\hat W_i^{(1)} =\p\rho_i + \Psi^{(h_i)}\t\Psi^{(h_i-1)}$
and $W_i^{(2)} =-[(\p\rho_i)^2-\p^2\rho_i + \Psi^{(h_i)}\p\t\Psi^{(h_i-1)}]$
Notice that  unlike the discussion above here there is only one scalar for
each system ($\rho_i$) rather than two ($\phi,\ \bphi$).
 Nonetheless, the transformations of the fields  $\Psi^{(h_i)},\
\t\Psi^{(1-h_i)}$ and $\rho_i$, which are given by eqn.
\mishbn\  and \mishbr\  with some
obvious renaming, leave the action  of eqn.
\mishbss\ invariant due to
the factor half in front of $\p\rho_i\bp\rho_i$.

We want to consider now the case of a non-flat world-sheet. Once we turn on
the curvature, the parameters $\alpha$ in the model of \mishab\  as well as $Q$
of the above discussion play an important role. In the case of the twisted
$N=2$ theory the level $k=({1\over \alpha^2}-2)$
 determines the dimension of the (moduli) space
on which all ``physical" correlators are cohomologies\refmark{\li,\dvv}.
Do we loose the symmetry structure generated by the infinitely many
generators
of eqn. \mishbs\  ? It turns out that  those invariances persist also
in the $R^{(2)}\ne 0$ case.  To realize this phenomena we use again the
 conformal metric $d s^2 =e^{\varphi}
dzd\bar z$. In this picture the action of eqn. \mishaa\  takes the form
 $$ \eqalign{S_{(N=2)}=&\int d^2 z \p\phi\bp\bphi +i\alpha\p\bp\varphi\phi
+(\l\p\bl +c.c)\cr
 =&\int d^2 z \p\phi\bp\hat\phi  +(\l\p\bl
+c.c) \qquad  \hat\phi =\bphi-i\alpha\varphi\cr}
\eqn\mishbr$$
This redefinition make sense if $\alpha$ is pure imaginary namely
for negative $k$. The later are natural if the starting point of the
twisted $N=2$ is $SL(2,R)$ WZW model rather than an $SU(2)$ model.
Following this redefinition  the form of all the generators remains the
same apart form the fact that $\hat\phi$ is replacing $\bphi$.

Two remarks are
in order: (i) Since a fixed world sheet metric was introduced it is clear
that the ghost sectors of the pure gravity theory have to be invoked and
hence the whole action of eqn. \mishaab\
 has to be added. The other remark refers
to the form of $T$. Building it from $G$ we get now
$$\oint_\w dz Q(z)  G(\w) =T(\w) = \p\phi(\p\bphi + i\alpha\p\varphi)+\l\bp\bl
-\p^2\phi \eqn\mishbs$$
This may look an unfamiliar expression but in fact this is exactly what has
to be achieved for this metric.\refmark\li
\section{ The algebraic structure}
\par An algebra for the TCFT's was written down in ref. [\dvv].
This algebra can be
deduced from the OPEs' of the various pairs of operators made out of
$J, T, Q$ and $ G$, using Jacobi identities. The OPEs' follow from the
``topological condition"
  given in eqn. \mishad\ . The algebra is characterized by
the  three anomalous terms in the Kac- Moddy algebra of $J$, in $[L,J]$ and in
$\{ G, Q\}$, which are all determined by one parameter denoted
in ref. [\dvv] as $d=d_{JJ}=d_{QG}=-d_{TJ}$.

Next we want to analyze the algebraic structure of the set $W^{(n)}, \Q $ and
$\G$. We first wish to confirm that the OPEs' which led to the algebra
of ref. [\dvv] are those of the symmetry generators for $n=1$ and $W^{(2)}$.
Using the definitions of eqns. \mishbm\ and \mishbp\ and the basic OPEs'
\mishbc\
it  is straightforward to check
that the resulting OPE's  :
$$ \eqalign{&W^{(1)}(z) Q^{(1)}(\w) = {Q^{(1)}(\w)\over(z-\w)}
\qquad W^{(1)}(z) G^{(1)}(\w) =- {G^{(1)}(\w)\over(z-\w)}\cr
&W^{(1)}(z) W^{(1)}(\w) = -{1\over(z-\w)^2}\qquad
W^{(1)}(z) W^{(2)}(\w) = {-1\over(z-\w)^3}-{W^{(1)}(\w)\over(z-\w)^2}\cr
&W^{(2)}(z) Q^{(1)}(\w) = {Q^{(1)}(\w)\over(z-\w)^2} +{\p
Q^{(1)}(\w)\over(z-\w)
}
\qquad W^{(2)}(z) G^{(1)}(\w) = {2G^{(1)}(\w)\over(z-\w)^2}
+{\p G^{(1)}(\w)\over(z-\w)}
\cr
&W^{(2)}(z) W^{(2)}(\w) =  {2W^{(2)}(\w)\over(z-\w)^2}
+{\p W^{(2)}(\w)\over(z-\w)}\cr
&Q^{(1)}(z) G^{(1)}(\w) = {-1\over(z-\w)^3}+{W^{(1)}(\w)\over(z-\w)^2}
 +{W^{(2)}(\w)\over(z-\w)}\cr}
\eqn\mishdaa $$
is identical to those of ref. [\dvv].
Since when acting on $\Q$ with $W^{(1)}$
it is in fact only $\hat W^{(1)}$
which operates, one can use the later as the ``ghost number current"
when acting on $\Q$.
Similarly one can use   $\hat{\t W^{(1)}}$ when applied on $\G$.
We now return to the more general form of the symmetry generators namely
those  with $\t a=-  a\ne 1$ given in eqn. \mishbd\ and in the discussion
following eqn. \mishbm\ .
It is straightforward to check that for this case one derives the same OPEs'
apart from the fact that now
$d=d_{JJ}=d_{QG}=-d_{TJ}= 2a\t a + 1$
For the parametrization of ref. [\li] one thus gets $d={k\over k+2}$.
Switching on $R^{(2)}$ introduces, as was explained in section 3, the
redefinition of $\t W\rightarrow \t W +i\alpha\p \varphi$.
It is easy to check that the OPEs' of eqn.
\mishdaa\ stay in tack under the
this modification.

We proceed now to the operators beyond the
``minimal topological algebra"\refmark\dvv.
First we examine  the OPE of  $W^{(1)}$ and
 $W^{(2)}$ with the rest of the operators.
In Appendix C it is shown in the context of model \mishaa\ that
$$ W^{(1)}(z) \Q(\w) = {\Q(\w)\over(z-\w)}
\qquad W^{(1)}(z) \G(\w) = {-\G(\w)\over(z-\w)}
\eqn\mishda $$
$${1\over 2\pi i}\oint_\w dz W^{(1)}(z)W^{(n)}(\w)=0 \eqn\mishdda$$
It is thus clear that  $W^{(1)}$ plays the role of the ghost number current
and that the $\Q$ and $\G$  have ghost number 1, -1 respectively.
 It is shown in Appendix C that the term proportional to
${1\over(z-\w)}$ in  $W^{(1)}(z) \W(\w)$ vanishes which leads to
 eqn. \mishdda\ .
 Hence, as expected from the
 its definition, $\W$ has a zero ghost number.
Similarly it is not surprising to notice that $W^{(2)}$
is the energy momentum tensor and $\Q, \ \G$ and $\W $ all carry
dimension equal to $n$. and $n+1$ respectively.

$$ W^{(2)}(z) \Q(\w) = ...{n(n-1)Q^{(n-1)}(\w)\over(z-\w)^3}
+{n\Q(\w)\over(z-\w)^2}  +{\p\Q(\w)\over(z-\w)}\eqn\mishdb$$
and similarly for $W^{(2)}\G$.

The next question of interest is whether the  OPEs' and the corresponding
commutation relations are linear or whether products of generators
and their derivatives
show up in them.
It turns out  that the algebra is not linear.
We demonstrate it now
in the following two examples:
$$ W^{(1)}(z) W^{(3)}(\w) = {-2\over(z-\w)^4}
+{2W^{(1)}(\w)\over(z-\w)^3}
-{[2W^{(2)}+:\t R^{(1)}\t R^{(1)}:+ \p\t R^{(1)}](\w)\over(z-\w)^2}
\eqn\mishddb$$
where the $: \ : $ denotes normal ordering as explained in Appendix C.
A similar structure show up in

$$Q^{(2)}(z) G^{(1)}(\w) = {-4\over(z-\w)^4}-{2W^{(1)}(\w)\over(z-\w)^3}
-{[2W^{(2)}+:\t R^{(1)}\t R^{(1)}:+ \p\t R^{(1)}](\w)\over(z-\w)^2}
 +{\Delta^{(3)}(\w)\over(z-\w)}
\eqn\mishddaa $$
Where $\Delta^{(3)}= -W^2\t W +\p(W^2) +\p W \t W +\p^2\bl \l +
2(\p W)\bl\l +2 W \p \bl \l $.
%We further present there
%the complete OPE of  $W^{(1)}(z) W^{(3)}(\w)$ to demonstrate that it is
%expressed in terms of normal ordered products of lower $\W$ and their
%derivatives.

Another obvious property of $Q^{(n)}$  and $G^{(n)}$ is nillpotency . This
is a special case of the anticommuting relations
 $$ \{ Q^{(n)} Q^{(m)}\}=0 \qquad \{ G^{(n)}, G^{(m)}\} =0 .\eqn\mishbr$$

The derivation of  $ Q^{(n)}(z) G^{(m)}(\w) $ is straightforward
 though tedious. In Appendix B
we present the calculation of the anomalous term.

\section {Summary and Discussion}
 Since the original path-integral approach  to TQFT's it was
known  that a basic
property of all the TQFTs' is the fact that all the non-zero
modes are canceled out from the ``physical observables".  This
characteristic  feature should manifest itself in terms of a large
set of symmetry constraints on physical states.
 In this note we have investigated the symmetry structure of
    topological theories.
We showed that the TCFTs' are in fact invariant under transformations
generated
by nilpotent pairs of fermionic operators of arbitrary conformal
dimension. An interesting feature of these generators is that they are
in fact the $n^{th}$ covariant derivative on the basic fermions of the
theory. The covariant derivative is with respect to a ``flat abelian
gauge connection". We showed that the generic model can be
derived as a BRST gauge fixed action of a theory  with a ``topological
symmetry" in which holomophicity condition was imposed.
It is thus plausible that the later construction and the infinite tower
of symmetries are related. In this case, it is not hard to envision, that
all the TCFTs' models that we considered are described by cohomologies
on moduli spaces of flat connections and their generalization to higher
spin fields.
The bosonic counterparts of the fermionic symmetries $\W$ and $\R$ where
also expressed as covariant derivatives.
The complete algebraic structure was not extracted in the present work.
Therefore it is not clear to what extent the algebra of the bosonic
generators is related to various $W_\infty$ which were discussed in the
literature.\refmark\Walgebra.
The implications of this very rich algebraic structure on
the Hilbert space of physical states is under investigation. We believe
that it is this algebraic structure which is responsible for the
decoupling of all the non-zero modes from the physical observables.
\par We did not discuss in this work the application of the ``minimal
toplogical algebra" to string theories. It was
realized\refmark{\dvv,\pol} that the set of $J,Q,G$ and $T$ do not close
the algebra and one has to introduce additional symmetry generators. It
was also found out that the non-critical string theory of $c=1$ share a
``higher symmetry".\refmark\roger
 The role of the new symmetries presented in this
work in the realm of string theories is under current investigation.

\ack{
J.S wants to thank R.D. Peccei and
the HET group of UCLA for the worm hospitality
during his stay  in UCLA where part
 of this work was done. He also wants  to thank R. Brooks for his
comments on the manuscript.
S.Y wants to thank the theory group of CERN.
Part of this work was done during his stay in CERN.}
\refout

\Appendix {A}
\par We want to show now that the $Q_\infty, G_\infty$ and $W_\infty$
transformations leave the TCFT models invariant. Again we present the
explicit proof for the $\alpha=0$ case of eqn. \mishbd\  and later we explain
how to extend the proof to the rest of the cases.
Obviously only $\l$ and $\bphi$  transform   by $\Q$. Recall  eqn.
\mishbn\
$$\delta^{Q^{(n)}}\l =  D_-^{n} \e \qquad
\delta^{Q^{(n)}}\bphi =
-\sum_{i=0}^{n-1} D_-^{i}\e D^{(n-1-i)}\bl. \eqn\mishAa$$
The action thus transforms into
$$\delta S_{N=2} =\int d^2 z[
\bp W\sum_{i=0}^{n-1} D^i\bl D_-^{(n-1-i)}\e + D_-^n\e\bp\bl]. \eqn \mishAb$$
Now this last expression is in fact a total derivative of the form
$\bp (D_-^n\e \bl) $. In order to prove that we have to show that
$$\bp W\sum_{i=0}^{n-1} D^i\l D_-^{(n-1-i)}\e = \bp (D_-^n \e )\bl
\eqn\mishAc$$
Let us expand the term on the right of the last equation:
$$ \bp (D^n \e )\bl =\bp (-\p + W) (D^{n-1}\e) \bl= \bp W (D^{n-1}\e) \bl +
\bp (D^{n-1} \e ) D\bl \eqn\mishAd$$
The term on the right is the first term in the sum of eqn. \mishAa\ . Further
iteration of expanding  the term to the right generates exactly all the terms
in the sum of eqn. \mishAd\ .
\par The invariance of the action under $\G$ follows from an identical proof
with the obvious replacements $\l \rightarrow \bl$ $W\leftrightarrow -\t W$

Now the generalization to the rest of the TCFT models is very straightforward.
For the $R^{(2)}\ne 0$ case one can again pass to the modified field
$\hat \phi$. As for the general case of eqn. \mishabb .
The same reasoning of
above leads to the conclusion that the variation of the action under for
example $\Q_i$ is $\delta S = \int d^2 z \bp (D_i^n \e \t \Psi^{(1-h_i)}$.

\par Next we want to explain
the relations of diagram \mishbe\ .
We use again the
example of \mishbd\ for $R^{(2)}=0$.
Let us show first  that
$$\Q =\oint_\w Q(z) \t \W(\w)= \oint W\bl(z)[ -D^{n-1}\t W - D^{n-1}(\bl \l)+
D^{n-1}\bl)\l] \eqn\mishAe$$
The first term in the integral gives $D^{n-1}(\p\bl)$. Plugging the OPE of
$\l \bl$ into the other two terms, recalling that $D\l=\p \l$ one gets
for the second term
$D^{n-1}(W\bl)$ so that altogether we get for the first two terms
$D^{n-1}(\p\bl + W \bl)= D^n\bl =
\Q$.  It is thus clear that omitting the last term $Q^{(n-1)}\l$ produces
$\R$.
Under the interchange of $\l$ with $\bl$ and $W $ with $-\t W$ we find
in a complete analogy the same results for $\G$.
%$$\Q =\oint_\w G(z) \W(\w)= \oint \t W\l(z)[ -\t D^{n-1}\ W+\t D^{n-1}(\l
% \bl)-

%\t D^{n-1}\l)\bl] \eqn\mishAf$$

Our next task to examine whether $\R$ generate symmetry transformations,
namely,
we want to check  if $\delta^{\R} S =
{1\over 2\pi i}\oint_\w dz [\e R^{(n)}](z) S =0 $.
Since we know that $\t R^{(n+1)}=\t W^{(n+1)}+\l\Q$ and since we know that
$\t \W$ are symmetry generators it is enough to show that $\l\Q$ leaves
the action invariant. It is straightforward to realize that the later holds.
The transformation of the various fields are found to be
$$\eqalign{&\delta^{R^{(n)}}\l =  -(\t D_-^{(n-1)} \e)\l \qquad
\qquad
\delta^{R{(n)}}\bl =  \t D_-^{(n-1)}( \e)\bl  \cr
&\delta^{R^{(n)}}\bphi = -\t D_-^{(n-1)}\e \qquad  \delta^{R^{(n)}}\phi=
-\sum_{i=0}^{n-2} \t D_-^{i}\e D^{(n-2-i)} \hat W^{(1)} -\t D_-^i (\e )
\t D^{(n-2-i)}(\l\bl) \cr}
 \eqn\mishAr$$

\Appendix {B}

We  compute  the
anomaly term in the OPE of  $Q^{(m)}G^{(n)}$. The notion of anomaly refers
here to the term proportional to ${1\over (z-\w)^{m+n+1}}$ which obviously
is a number.
 One gets this term by performing a
 complete contraction of all the fields. For $n>m$
the general form of a term in the expansion which can contribute to the
anomalous term has the form
$$[W^i\p^j W \p^{m-(i+j)-1}\bl](z)[\t W^i\p^l \t W \p^{n-(i+l)-1}\l](\w)
\eqn\mishBa$$
where $i=0,...m-1$, $j=1,...m-(i+1)$ and $l=1,.... n-(i+1)$. In addition
one can have the case with no derivatives on $W$ and $\t W$.
The contribution
of a term of the form of \mishBa\  is found by performing all possible
contractions  between the fields. One gets
$$(-1)^m [m+n-(2i+j+l)]!i!(j+1)!(l+1)!+(l+j+1)!]\eqn\mishBb$$
The contribution of the terms with no $W$ and $\t W$ derivatives are
$(-1)^m[m+n-2i]!i!$. What  is left over to do is to figure out the
multiplicity factors $B_i$ and $D_{ijl}$
of each of the terms and then perform the summation, namely:
$$ \eqalign{Anom =&(-1)^m\sum_{i=0}^m B_i[(m+n)-2i]!i!\cr
&(-1)^m\sum_{i=0}^m-1\sum_{j=1}^{m-i-1} D_{ijl}[(m+n)-2(i+1)-(j+l)]!i!
[(j+1)!(l+1)!+(l+j+1)!]\cr}\eqn\mishBc$$
It is easy to check that $B_i = (^m_i )(^n_i )$  and  similarly one can
get an expression for $D_{ijl}$

\Appendix {C}

As in the previous sections we work here in the context of the flat world
sheet of eqn. \mishaa\ . Thus following eqn. \mishbp\ $W^{(1)} = (W+\t W
-\l\bl)$. When acting on $\Q =D^n\bl$ obviously only the second and the
third terms in $W^{(1)}$ can contribute. Let us first look on the residue,
namely, the ${1\over (z-\w)}$ terms. Since following eqn. \mishbc\ the OPE
$\t W(z) W(\w)={1\over (z-\w)^2}$ and when $\t W$ is
applied on $\Q(\w)$ there are
no terms at $z$ to expand, the only contributions can come from
$\l\bl(z) \Q(\w)$. Denoting a generic term in $\Q$ as
$C_k F_k(W,\p W) \p^{n-k}\bl$ where $C_k$ is some numerical coefficient and
$F_k(W \p W)$ is a dim $k$ polynomial of $W$ and derivatives of $ W$, than

$$ \l\bl (z) C_k F_k \p^{n-k}\bl (\w)=-C_k F_k \bl (z)
{[n-k]! \over (z-\w)^{n+1-k}}
=...-{1\over (z-\w)}C_k F_k(W,\p W) \p^{n-k}\bl \eqn\mishCa$$
It is thus clear that the residue is really $-\Q$.

We want to show now that
all the terms multiplying ${1\over (z-\w)^l}$ for $l>1$ vanish.
Terms proportional to ${1\over (z-\w)^2}$ are generated by contraction between
the $\l\bl$ and $\Q$ and between $\t W$ and powers of W in $f_k$. Rewriting
the later as $C_k f_k =C_{i,k}W^i g_{k-i}$ we get a contribution of
$-\sum_{i,k}C_{i,k}W^i(n-k)\p^{n-k}\bl$ where as the $\t W W$ contractions
lead
$\sum_{i,k}C_{i,k}iW^{i-1}(n-k)\p^{n-k}\bl$. Now since
$(i+1)C_{i+1,k}= (n-k)C_{i,k}$ for $k\ne i+1$ and
$(i+1)C_{i+1,k}= (n-i)C_{i,k}$
 for $k=i+1$ the two contributions cancel each
other.

Next we compute the terms multiplying ${1\over (z-\w)^j}$ for $j=1,2,3$.
Following the same steps as for $W^{(1)}$ one can realize that from
contraction the $\l\p\bl (z)$ term one get
$-C_k F_k {(n-k)!\over l!(z-\w)^{n+1-k-l}}\p^{l+1}\bl $. For
$l=n-k$ one can exactly the action of the derivative on $\bl$ in $\p \Q$.
When $\t W W$ is contracted with powers of $W$ in $f_k$ one gets the
action of the derivative of this part and same applies for $\p^j W$ factors
in $f_k$. So altogether applying the chain rule one gets the term
${\p\Q\over (z-\w)}$. Repeating the analysis now for $j=2,3$ one derives the
eqn. \mishdb\ .
\par We present here the explicit calculation of $W^{(1)} W^{(3)}$
 The terms multipying  the various powers of
 ${1\over (z-\w)}$ are
 $$\eqalign{{1\over (z-\w)^3} &:
 4\t W + 2W -2\l\bl -2\t W =2 W^{(1)}\cr
  {1\over (z-\w)^2} &:
2\p W -\l\p\bl +\p\t W -2 W \t W -\t W^2 + 2\bl\p\l -2\bl W\bl\cr
& -[W^{(2)} +
:R^{(1)}\t R^{(1)}:
+ \p \t R^{(1)}]\cr}\eqn \mishCF$$
where the normal order product
$:R^{(1)}\t R^{(1)}:$ is given by $W^2 +2W\l\bl + \p\l\bl +\p\bl\l$.
\end